\documentclass[letterpaper,prl,twocolumn,showpacs,superscriptaddress,amsmath,amssymb]{revtex4}

\usepackage{graphicx}
\usepackage{dcolumn}
\usepackage[latin1]{inputenc}
\usepackage{bm}

\usepackage{color}
\usepackage[dvipsnames]{xcolor}

\DeclareGraphicsExtensions{.eps} 

\begin{document}

\title{General Cramér-Rao bound for parameter estimation using Gaussian multimode quantum resources}

\author{O. Pinel}
\affiliation{Laboratoire Kastler Brossel, Université Pierre et Marie
Curie-Paris 6,\\
ENS, CNRS; 4 place Jussieu, 75252 Paris, France}
\author{J. Fade}
\affiliation{Institut Fresnel, CNRS, Aix-Marseille Université, Ecole Centrale Marseille, Campus de Saint-Jérôme,13013 Marseille, France}
\author{N. Treps}
\affiliation{Laboratoire Kastler Brossel, Université Pierre et Marie
Curie-Paris 6,\\
ENS, CNRS; 4 place Jussieu, 75252 Paris, France}
\author{C. Fabre}
\affiliation{Laboratoire Kastler Brossel, Université Pierre et Marie
Curie-Paris 6,\\
ENS, CNRS; 4 place Jussieu, 75252 Paris, France}

\date{\today}

\begin{abstract}
Multimode Gaussian quantum light, including multimode squeezed and/or multipartite quadrature entangled light,
is a very general and powerful quantum resource with promising applications to
quantum information processing and metrology involving continuous variables. In this paper, we determine the ultimate sensitivity in the estimation of any parameter when the information about this parameter is encoded in such Gaussian light, irrespective of the exact information extraction protocol used in the estimation. We then show that, for a given set of available quantum resources, the most economical way to maximize the sensitivity is to put the most squeezed state available in a well-defined light mode. This implies that it is not possible to take advantage of the existence of squeezed fluctuations in other modes, nor of quantum correlations and entanglement between different modes. We show that an appropriate homodyne detection scheme allows us to reach this Cramér-Rao bound. We apply finally these considerations to the problem of optimal phase estimation using interferometric techniques.

\end{abstract}

\maketitle


Optical techniques are widely used in many areas of science and technology to perform high performance
measurements and diagnostics, for example in spectroscopy, ranging,
trace detection, microscopy, image processing. There are many reasons for this: light
allows us to extract information in a remote and non destructive way,
it carries information in a massively parallel way, and perhaps more
importantly optical measurements can reach very high precision and/or
sensitivity levels, the best example being the interferometers used as
gravitational wave antennas. In this respect, it is very important to
know exactly and in the most general way what are the limits of
accuracy in parameter estimation using light. These limits depend on
the noise present in the detection process, which is ultimately, when
all sources of imperfections in the setup have been eliminated, the
quantum fluctuations of the photodetection signals used to estimate the
parameter of interest.

When the detected light is produced by ordinary, shot noise limited, sources, such a limit is called "standard quantum limit". It is well-known that it is possible to improve optical measurements beyond the standard quantum limit by using squeezed\cite{sq} or entangled\cite{en} light. This statement has been demonstrated for what can be called "simple measurements", in which the information about the parameter $p$ to measure is carried by the total intensity\cite{int} or by the phase\cite{pha,Caves} of a single-mode light beam. It has been recently extended to the case of optical images in which the parameter $p$ to measure does not change the total intensity of the light but instead modifies the details of the repartition of light in the transverse plane\cite{treps}, and experiments have been performed for small displacement measurements\cite{displacement}.

The purpose of this paper is to derive the expression of the ultimate sensitivity in the estimation of a parameter encoded in any kind of variation, spatial and/or temporal, of an optical field. To this purpose, we will determine the Cramér-Rao bound on the parameter estimation, which gives a limit to the best expectable precision, independently of the strategy used in the estimation \cite{ref04a,gar95,Helstrom}. In addition, we will consider that the light which is used in the measurement belongs to a wide set of quantum states, namely the set of multimode Gaussian states, which contains all possibilities of multimode squeezing as well as all kinds of multipartite quadrature entanglement. These non-classical states are now produced by experimentalists with impressive amounts of squeezing or entanglement \cite{Furusawa}. We will also assume that the information that is extracted from the light and then processed is the intensity, or a field quadrature, at different positions and/or times. Our approach does not give the most general "quantum Cramér-Rao bound"\cite{Giovanetti,Berry,Goldstein}, optimized over all possible quantum states and all possible quantum measurements, but instead the lowest possible uncertainty optimized over a subset of quantum states and quantum measurements that are readily available using present technology even for very large values of the mean photon number (up to $10^{16}$), which is not the case with more exotic non-Gaussian states and detectors like NOON states and Photon Number Resolving detectors.

\begin{figure}[htbp]\label{setup}
\centerline{\includegraphics[width=8cm]{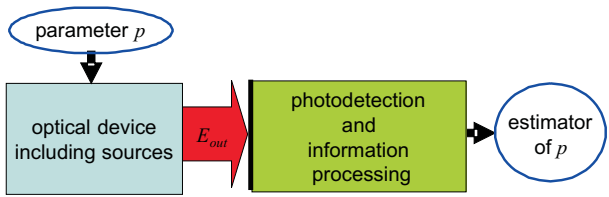}}
\caption{General scheme for estimating light parameters}
\label{sketch}
\end{figure}

A sketch of a general optical measurement setup is displayed in the figure \ref{sketch}. An optical device, containing light sources and various optical elements, produces an output beam of light described at the detector position by its complex electric field ${\bf E}_{\text{out}}$, which is multimode in space and time in the general case. The parameter that one wants to determine, called $p$, modifies in a known way the output beam, so that ${\bf E}_{\text{out}}={\bf E}_{\text{out}}(p)$. One then uses a photo-detector that records a $p$-dependent information. This information is a classical quantity which is processed by analog or digital techniques. The outcome is a real number which is used as an {\it estimator} of $p$, that we assume to be unbiased.

In the following, we assume that both the optical device and the photodetector are perfect, i.e. that all causes of {\it technical noise} have been eliminated. The only source of uncertainty in such a measurement is thus the unavoidable fluctuations in the photodetection signal arising from the quantum fluctuations of the light beam ${\bf E}_{\text{out}}$, and that can be tailored in the optical device producing the output beam in such a way that ${\bf E}_{\text{out}}$ may have either squeezed or quantum-correlated quadrature fluctuations, or both.

To simplify the formalism we will assume that the output field is
polarized in a fixed direction, so that one can use a scalar
description of the output field $E_{\text{out}}(p)$. In addition we will use
the paraxial and slowly varying envelope approximations when
necessary. We will also assume that the number of modes necessary to
describe the output field is finite, and has a value $M$. The mean output complex field can then be written
\begin{equation}
  \overline{E}_{\text{out}}({\bf r},t,p)=\mathrm{i}\sqrt{\frac{N \hbar \omega}{2 \varepsilon_0 c T}}u_1({\bf r},t,p)
\end{equation}
where $u_1(r,t,p)$ is a normalized mode that follows the evolution of
the mean field with $p$, $T$ the exposure time and $N$ the mean value of the total photon number. We will leave aside the already well known case where the parameter is encoded in the light intensity, and assume that $N$ is not modified by the parameter $p$.

We introduce a general mode basis to describe the detection process that we
call $\{v_i({\bf r},t)\}$ ($i=1,...,M$), which constitutes a complete
basis for the description of the electric field at the location of the
photodetector. One can write the positive
frequency electric field operator
\begin{equation}
\hat{E}^{(+)}_{\text{out}}({\bf r},t)=\mathrm{i}\sqrt{\frac{\hbar \omega}{2 \varepsilon_0 c T}}\sum_i \hat{b}_i v_i({\bf r},t)
\end{equation}
where $\hat{b}_i$ is the annihilation of photons operator in mode $v_i$. The
modes $v_i({\bf r},t)$ can be for example the Hermite-Gauss modes in
the pure spatial domain, or a set of Gaussian temporal modes if one
wants to describe light pulses of arbitrary shapes. We write $|\psi(p)\rangle$ the quantum state produced by the device and which depends on $p$. One has $\overline{E}_{\text{out}}({\bf
  r},t,p)=\langle\psi(p)|\hat{E}^{(+)}_{\text{out}}({\bf
  r},t)|\psi(p)\rangle$ (the bar denoting in all this paper the
quantum mean value). We also introduce the quadrature operators for mode $i$: $\hat Y_i^+ = \hat b_i^\dagger+\hat b_i$ and $\hat Y_i^- = i(\hat b_i^\dagger-\hat b_i$) and the general matrix column quadrature operator
$ \hat{\bf Y}=(\hat{Y}^+_1, ..., \hat{Y}^+_M,\hat{Y}^-_1, ..., \hat{Y}^-_M)^{\top} $.

We assume that $|\psi(p) \rangle$ is a Gaussian state of light. It can then be described by a Gaussian Wigner
function and its properties are entirely defined by the mean
value and covariance matrix of its quadratures operators. Let us call $\overline{\bf Y}(p)$
the mean value of the quadratures and $\bm{\Sigma}(p)$ the covariance matrix written in the basis $\{v_i, \hat b_i\}$, which contains all the information about the multimode entanglement and squeezing properties of the state. Its Wigner function is then expressed as
\begin{equation}\label{Wigner}
W_p({\bf Y})=\frac1{(2 \pi)^M} \exp\left(-\frac12({\bf Y}-\overline{{\bf Y}}(p))^{\top}\bm{\Sigma}^{-1}(p)({\bf Y}-\overline{{\bf Y}}(p))\right)
\end{equation}
In order to get simple expressions, we assume that we have a pure state ($\text{det}[ \bm{\Sigma}]=1$), but one can easily generalize the derivation of the Cramér-Rao bound to a mixed state.

Let us now calculate the Fisher information, inverse of the Cramér-Rao bound, corresponding to the detection of a variation of the parameter $p$ from its initial value, that one can always take as $p=0$ by a change of origin. We need to assume here for the calculation that one is able to measure the two quadratures of all the
modes of the field. Even if non-physical, as two quadratures of the
same mode correspond to non commuting observables, this assumption
will give an upper bound to the Fisher information as it increases the
information that one can actually extract from the field. More
physically, for a stationary situation where the same state of light is continuously produced by the device, it corresponds to measurements
performed during twice the time, once on each quadrature. This means that we have complete access to the vector $\mathbf{Y}$. In this case, and assuming that we have only unbiased estimators, the Fisher information has the following expression \cite{ref04a,gar95}:
\begin{equation*}
I_{\scriptscriptstyle\text{Fisher}}=-\int\frac{\partial^{2}l({\bf Y}|p)}{\partial p^{2}}\biggr|_{p=0}L({\bf Y}|p=0)\,d{\bf Y}
\end{equation*}
$L({\bf Y}|p)$ being the likelihood of measuring ${\bf Y}$ knowing that the parameter value is $p$ and $l({\bf Y}|p)$ being the log-likelihood.

Using the Wigner function defined in eq. \ref{Wigner} one finds that
\begin{eqnarray}\label{IF}
I_{\scriptscriptstyle\text{Fisher}} & = & \left(\frac{\partial\overline{{\bf Y}}}{\partial p}(p)\biggr|_{p=0}\right)^{\top} \bm{\Sigma}^{-1}(p=0)\left(\frac{\partial{\bf \overline{Y}}}{\partial p}(p)\biggr|_{p=0}\right)\nonumber \\
& & +\frac{1}{2}\text{Tr} \left[  \bm{\Sigma}(p=0) \frac{\partial^{2} \bm{\Sigma}^{-1}(p)}{\partial p^{2}} \biggr|_{p=0}\right]
\end{eqnarray}
This expression gives the Fisher information as a function of the field mean value and variance, and of its dependence with $p$. It is made of two terms. The first term is linked to the variation of the mean field distribution with $p$ while the noise taken into account is the one for $p=0$. The second term comes from the modification of the covariance matrix of the field with $p$ in the neighborhood of $0$, and it does not depend on the mean field, thus it does not scale with the mean number of photons $N$. For large value of $N$, as those involved in continuous variable regime, it becomes negligible compared to the first term of the Fisher information, and we will neglect it in the rest of this paper. In the remainder of this paper, we will also restrict our analysis to the case of the estimation of very small variations of the parameters value, where the effort towards precision estimation enhancement is most crucial.

We now choose an adequate detection basis in order to obtain an interpretable form of the Cramér-Rao bound. The idea is to find a basis where $\frac{\partial{\bf \overline{Y}}}{\partial p}\bigr|_{p=0}$ is single mode. Let us define the first mode of the detection basis by
\begin{equation}\label{det}
v_{1}=p_c \frac{\partial u_{1}(p)}{\partial p}\biggr|_{p=0} \quad \textrm{where } \frac{1}{p_c}=\left\Vert \frac{\partial u_{1}(p)}{\partial p}\biggr|_{p=0}\right\Vert
\end{equation}
and put no additional constraints on the other modes $v_{n>1}$ of the basis (except that they are orthogonal to $v_{1}$). Let us stress that this basis does not depend on the actual value of $p$. One has $\frac{\partial{\bf \overline{Y}}}{\partial p}\bigr|_{p=0}= 2\sqrt{N} v_1/p_c$. The calculation of the Fisher information in that basis is then very simple as only one vector is involved and thus only one element of $\bm{\Sigma}^{-1}(0)$.
\begin{equation}\label{F2}
I_{\scriptscriptstyle\text{Fisher}} = \frac{4N}{p_c^2}\bm{\Sigma}_{(1,1)}^{-1}(0)
 \end{equation}
where $\bm{\Sigma}_{(1,1)}^{-1}(0)$ is the first left, top element of
the matrix $\bm{\Sigma}^{-1}(p=0)$. We find that the Fisher information only depends on the mode $v_1$ defined by the derivative of the mean field mode with respect to $p$, that we will call from now on the \textit{detection mode}\cite{multipixel}.

We are then led to our final expression of the Cramér-Rao bound for unbiased parameter estimation using quantum Gaussian states:
\begin{equation}\label{F3}
\Delta p_{\scriptscriptstyle\text{CRB}} =\frac{1}{\sqrt{I_{\scriptscriptstyle\text{Fisher}}}}=\frac{p_c}{2\sqrt{N}}\frac1{\sqrt{\bm{\Sigma}_{(1,1)}^{-1}(0)}}.
\end{equation}
The Cramér-Rao bound we have just derived depends only on three parameters: a scaling factor $p_c$ that characterizes the variation of the mean field distribution with $p$, the mean total number of photons and one element of the inverse of the covariance matrix. In this article we take for given $p_c$ and $N$ as optimizing the mean field shape and increasing the number of photons are two well known strategies to increase information extraction and we assume here that they have both been optimized. We focus here on the effect of the noise, and the remarkable result that we have obtained is that the Cramér-Rao bound depends only on the diagonal matrix element of the inverse of covariance matrix on the detection mode.

The Cramér-Rao bound (\ref{F3}) gives the best estimation precision one can hope to obtain on a parameter encoded on a specific Gaussian quantum state, characterized by its covariance matrix. The question is now to find the best strategy to optimize this bound given a certain amount of quantum resources. Let us start, like in most experiments to date, from $s$ squeezed vacuum states. $\sigma_i$ ($i=1,...s$) is the r.m.s.~value of the squeezed quadrature of mode $i$, and we call $\sigma_{\text{min}}$ the smallest of its values.  With the help of linear couplers i.e.~of unitary transformations\cite{Braunstein} on the mode basis, the multimode squeezing can be transformed partially or totally into multipartite entanglement in a different mode basis. One can show that, under such unitary transformations, the diagonal coefficients of the inverse of the covariance matrix are bound by its spectral radius, which is equal to $1/\sigma_{\text{min}}^2$. The equality is reached only when the detection mode 1 is an eigenmode of the covariance matrix with the eigenvalue $\sigma_{\text{min}}^2$, and thus \textit{when the detection mode is the most squeezed mode and is not correlated with all the other modes}.  The optimized Cramér-Rao bound is thus
\begin{equation}\label{CRBopt}
\Delta p_{\text{opt}} =\frac{p_c}{2\sqrt{N}}\sigma_{\text{min}}.
\end{equation}

We have shown here that the only way to minimize the Cramér-Rao bound given a certain initial amount of quantum resources is to put the most squeezed state available in the detection mode and not to have correlations with other modes. The presence of other squeezed modes, or of any kind of entanglement, will not help to improve the limit: one cannot take advantage of squeezed fluctuations or quantum correlations coming from different modes to improve the estimation of a single parameter\cite{Tilma}.

The determination of the Cramér-Rao bound does not tell us how to reach it, or even whether it is actually possible to reach it. In the present case, our analysis has outlined the importance of what we have called the detection mode. We will now show that a balanced homodyne detection scheme in which the local oscillator is taken in the detection mode allows us to reach the Cramér-Rao bound.

When $p$ is close enough to its initial value $0$, the mean field mode $u_1(p)$ can be expanded as
\begin{equation}\label{Taylor}
 u_{1}(p)\approx u_{1}(0)+p\frac{\partial u_{1}}{\partial p}\biggr|_{p=0} = u_1(0) + \frac{p}{p_c} v_1
\end{equation}
For a small non-zero value of $p$, the field is therefore composed of the original field plus the detection mode $v_1$ times the parameter.
In the general case, $v_1$ is not orthogonal to $u_1(0)$. We define a detection orthonormal basis whose first mode is $v_1$ and second mode, in the subspace spanned by $\{u_1(0), v_1\}$, is $v_{2}=\frac{u_{1}(0)-(u_{1}(0).v_{1})v_{1}}{\left\Vert u_{1}(0)-(u_{1}(0).v_{1})v_{1}\right\Vert }$. Noting that $\intop u_{1}^{*}(0)\frac{\partial u_{1}}{\partial p}|_{p=0}$ is an imaginary number, we can write $u_1(0) = ic_{11}v_1 + c_{12}v_2$ where $c_{11}$ is a real number. Finally, $u_1(p)$ can be expressed in the detection basis as
\begin{equation}
u_{1}(p)=\left( \frac{p}{p_c} + \mathrm{i}c_{11}\right)v_{1}+c_{12} v_{2}
\end{equation}

On can show that\cite{Delaubert06}, using a homodyne detection scheme in which the local oscillator is a coherent state in the $v_1$ mode, and if the relative phase between the field to detect and the local oscillator is zero, the intensity difference operator between the two outputs of the set-up is given by
\begin{equation}
\hat{I}_-=\frac{\hbar\omega}{2\varepsilon_0cT}\sqrt{N_{0}}\left(2\sqrt{N}\frac{p}{p_c} + \delta\hat Y_1^+\right)
\end{equation}
where $N_0$ is the local oscillator mean number of photon and $\delta\hat Y_1^+= \hat Y_1^+ - \langle\hat Y_1^+\rangle$ is the quantum noise fluctuation operator of the incoming field in $v_1$ mode. From that expression, it is easy to derive both the signal $\langle \hat{I}_-\rangle$ and the noise $\sigma_{\hat I_-}$. The $p$ value giving a signal to noise ratio equal to 1 is
\begin{equation}
p_{\scriptscriptstyle{\text{SQL}}}=\frac{\sigma_{Y_{1}^{+}}}{2\sqrt{N}\left\Vert \frac{\partial u_{1}}{\partial p}\bigr|_{p=0}\right\Vert }
\end{equation}
which is nothing else than the Cramér-Rao bound (\ref{CRBopt})

We will finally illustrate the interest of our approach by revisiting a well-known and important problem of quantum optics, namely the interferometric measurement of a phase shift  $\phi$ \cite{pezze,Higgins,Dorner,Anisimov}. As a Michelson interferometer has two output modes $v_1$ and $v_2$ than can be detected separately, it is a two-mode problem. In presence of a phase-shift $\phi$ between the two arms of the interferometer close to a bias value $\phi_0$, the total mean output field is
\begin{equation}\label{mean}
u_1(\phi)=v_1 \cos(F(\phi/2)) + v_2 \sin(F(\phi/2))
\end{equation}
When the arms are empty, $F(x)=x$, but optical devices, such as Fabry-Perot cavities, can be inserted in the two paths in order to increase the phase sensitivity of the interferometer. A complete basis of our two-mode space is made of the mode $u_1$ completed by the orthogonal mode $u_2(\phi)=- v_1 \sin(F(\phi/2)) + v_2 \cos(F(\phi/2))$. The detection mode defined in (\ref{det}) is in the present case $u_2(\phi_0)$, and the Cramér-Rao bound, according to equation(\ref{CRBopt}), is
\begin{equation}
\Delta \phi=\frac1{|F'|\sqrt{N \bm{\Sigma}^{-1}}_{11}}
\end{equation}
It is independent of the initial phase bias $\phi_0$, and is minimum when $F'$ is maximum, as could be expected. Its minimum value is obtained by having at the output of the interferometer a quantum state consisting of a squeezed state in the detection mode $u_2(\phi_0)$ and any Gaussian state in the orthogonal mode $u_1(\phi_0)$ uncorrelated with the first one. It is easy to see that such an output state is obtained by sending at the two input ports of the interferometer a tensor product of a field with nonzero mean value and of a vacuum squeezed state: one thus finds that the well-known scheme introduced by C. Caves\cite{Caves} is optimal when one uses Gaussian resources. This implies in particular that the use of Gaussian entangled states of the two input modes will not help improve the sensitivity of the interferometer\cite{Tilma}.

Let us finally mention that these results can be generalized to the estimation of several parameters: one can show that when these parameters are "orthogonal", it is possible to optimally reduce the noise on their estimators at once by independently squeezing the two corresponding detection modes. We now plan to extend the present approach to various kinds of non-Gaussian states and detectors in the continuous variable regime, such as Schrödinger cats and multiphoton detectors.

We acknowledge the financial
support of the Future and Emerging Technologies (FET) programme within
the Seventh Framework Programme for Research of the European
Commission, under the FET-Open grant agreement HIDEAS, number
FP7-ICT-221906


\begin{thebibliography}{99}

\bibitem{sq} H.A. Bachor {\it et al.}, {\it A guide to experiments
in quantum optics}, Wiley-VCH (2003).

\bibitem{en} P. Kok {\it et al.}, J. Opt. B {\bf 6}, S811 (2004).

\bibitem{int} F. Marin {\it et al.}, Opt. Comm. {\bf 140}, 146 (1997).

\bibitem{pha} M. Xiao {\it et al.}, Phys. Rev. Lett. {\bf 59}, 278 (1987).

\bibitem{Caves} C. M. Caves, Phys. Rev. D {\bf 23}, 1693 (1981).

\bibitem{treps} V. Delaubert {\it et al.},  Europhys Lett  {\bf 81}  44001 (2008).

\bibitem{displacement} N. Treps {\it et al.}, Science, {\bf 301}, 940 (2003).

\bibitem{ref04a} P. R\'efr\'egier {\it Noise Theory and Application to Physics} Springer, New-York (2004).

\bibitem{gar95}P.H. Garthwaite {\it et al.}, {\it Statistical Inference} Prentice Hall Europe, London (1995).

\bibitem{Helstrom} C.W. Helstrom, IEEE Trans. Information theory, {\bf 14}, 234 (1968).

\bibitem{Furusawa} M. Yukawa {\it et al.}, Phys. Rev. A {\bf 78}, 012301 (2008).

\bibitem{Giovanetti} V. Giovannetti {\it et al.}, Phys. Rev. Lett. {\bf 96}, 010401 (2006).

\bibitem{Berry} D.W. Berry {\it et al.}, Phys. Rev. A {\bf 80}, 0052114 (2009).

\bibitem{Goldstein} G. Goldstein {\it et al.} arXiv:1001.4804v [quant-ph] (2010).

\bibitem{multipixel} N. Treps {\it et al.}, Phys Rev A {\bf 71} 013820 (2005).

\bibitem{Braunstein} S.L. Braunstein, Phys. Rev. A {\bf 71} 055801 (2005).

\bibitem{Tilma} T. Tilma {\it et al.}, Phys. Rev. A {\bf 81}, 022108 (2010).

\bibitem{Delaubert06} V. Delaubert {\it et al.},  Phys Rev A {\bf 74} 053823 (2006).

\bibitem{pezze} L. Pezze {\it et al.}, Phys. Rev. Letters {\bf 100} 073601 (2008).

\bibitem{Higgins} B. Higgins {\it et al.}, Nature {\bf 450} 393 (2007).

\bibitem{Dorner} U. Dorner {\it et al.}, Phys. Rev. Letters {\bf 102} 040403 (2009).

\bibitem{Anisimov} P. Anisimov {\it et al.}, Phys. Rev. Lett. {\bf 104}, 103602 (2010).



\end{thebibliography}
\end{document}